\documentclass[amsmath,amssymb,12pt,superscriptaddress,nofootinbib]{revtex4}

\usepackage[dvipdfmx]{graphicx}
\usepackage{dcolumn}
\usepackage{bm}
\usepackage{color}

\newcommand{\dd}{\mathrm{d}}

\newcommand{\del}{\partial}
\newcommand{\ee}{{\rm e}}

\definecolor{DarkBlue}{rgb}{0,0,0.7} 

\definecolor{DarkRed}{rgb}{0.65,0,0}

\begin{document}
\baselineskip5.5mm

\thispagestyle{empty}

{\baselineskip0pt
\small
\leftline{\baselineskip12pt\sl\vbox to0pt{
               \hbox{\it Division of Particle and Astrophysical Science, Nagoya University}
               \hbox{\it Department of Physics, Rikkyo University} 
		\hbox{\it Yukawa Institute for Theoretical Physics, Waseda University} 
               \hbox{\it Advanced Research Institute for Science and Engineering, Waseda University} 
                             \vss}}
\rightline{\baselineskip16pt\rm\vbox to20pt{
            {
            \hbox{RUP-16-31}
            }
\vss}}%
}

\author{Chul-Moon~Yoo}\email{yoo@gravity.phys.nagoya-u.ac.jp}
\affiliation{
Division of Particle and Astrophysical Science,
Graduate School of Science, Nagoya University, 
Nagoya 464-8602, Japan
}

\author{Tomohiro~Harada}\email{harada@rikkyo.ac.jp}
\affiliation{
Department of Physics, Rikkyo University, Toshima,
Tokyo 171-8501, Japan
}

\author{Hirotada~Okawa}\email{h.okawa@aoni.waseda.jp}

\affiliation{Yukawa Institute for Theoretical Physics,
Kyoto University, Kyoto 606-8502, Japan
}
\affiliation{
Advanced Research Institute for Science and Engineering, 
Waseda University, 3-4-1 Okubo, Shinjuku, 
Tokyo 171-8501, Japan
}

\vskip1cm
\title{3D Simulation of Spindle Gravitational Collapse 
\\ of a Collisionless Particle System}

\begin{abstract}
We simulate the 
spindle gravitational collapse of a collisionless particle system
in a 3D numerical relativity code 
and compare the qualitative results with the old work done 
by Shapiro and Teukolsky~\cite{Shapiro:1991zza}. 
The simulation starts from the prolate-shaped distribution of
 particles and a spindle collapse is observed. 
The peak value and its spatial position of curvature invariants 
are monitored during the time evolution. 
We find that the peak value of the Kretschmann invariant
takes a maximum at some moment, when there is no apparent horizon, 
and its value is greater for a finer resolution, which is consistent
 with what is reported in Ref.~\cite{Shapiro:1991zza}. 
We also find a similar tendency for the Weyl curvature invariant. 
Therefore, our results lend support to the 
formation of a naked singularity 
as a result of the axially symmetric spindle collapse of 
a collisionless particle system 
in the limit of infinite resolution. 
However, unlike in Ref.~\cite{Shapiro:1991zza}, 
our code does not break down then but go well beyond.
We find that the peak values of the 
curvature invariants start to 
gradually decrease with time for a certain period of time. 
Another notable difference from Ref.~\cite{Shapiro:1991zza} is 
that, in our case, the peak position of the Kretschmann curvature invariant 
is always inside the matter distribution. 
\end{abstract}

\maketitle
\pagebreak

%
\section{Introduction}

Gravitational collapse is one of the most typical and attractive phenomena 
in general relativity. 
The singularity theorem (see, e.g., Ref.~\cite{Hawking:1973uf})
states that the formation of spacetime singularities is inevitable 
as a result of gravitational collapse with physically reasonable matter fields. 
If the cosmic censorship conjecture, 
proposed by Penrose~\cite{1969NCimR...1..252P,2002GReGr..34.1141P,Penrose1979}, 
is valid, those singularities generated from general and physically reasonable 
initial data should be clothed by a black hole horizon. 
Visible spacetime singularities are so-called naked singularities and 
a bunch of examples for a naked singularity are reported in various spacetimes. 
Generality of naked singularity formation is an important open issue 
in general relativity.

The cosmic censorship conjecture and naked singularity formation 
is of interest not only in a mathematical aspect of general relativity, 
but also in finding the cut-off energy scale of general relativity. 
In other words, 
a spacetime domain near a naked singularity with infinite curvature, 
which we call a border of spacetime~\cite{Harada:2004mv}, 
may be a window into new physics beyond general relativity. 
Even if the curvature scale does not exceed the cut-off scale, 
it would provide the locally high energy region 
in which unknown high energy particle physics phenomena may take place. 
The higher curvature regions associated with gravitational collapse
would provide a key to understand unsolved problems in 
cosmology, astrophysics and high energy particle physics.

In this paper, we focus on non-spherical gravitational collapse. 
As is stated by the hoop conjecture~\cite{Thorne1972}, 
a gravitational source with a sufficiently large 
circumference compared to its 
gravitational radius cannot form a black hole with a horizon. 
Therefore, we expect that the gravitational collapse which causes a 
highly elongated or flattened object at the end may not be 
surrounded by a black hole horizon and produces a spacetime border. 
One of the most famous examples has been presented 
by Shapiro and Teukolsky~(ST)~\cite{Shapiro:1991zza}, 
where the violation of the cosmic censorship conjecture 
due to spindle gravitational collapse of collisionless 
ring sources is discussed assuming axisymmetry. 
Our purpose in this letter is the reanalysis of this system 
by using recently developed numerical relativity techniques without exact axisymmetry.

Shapiro and Teukolsky~\cite{Shapiro:1991zza} firstly dealt with 
relativistic collisionless matter in axisymmetric 
spacetimes (see Ref.~\cite{Yamada:2011br} for a higher-dimensional version). 
Full 3-dimensional simulations of relativistic collisionless particle systems 
have been performed by Shibata in Refs.~\cite{Shibata:1999,Shibata:1999wi}. 
We basically follow the methods adopted in Refs.~\cite{Shibata:1999,Shibata:1999wi}. 
The specifications of our numerical procedure are described in Sec.~\ref{method}. 

In this paper, we use the geometrized units in which both
the speed of light and Newton's gravitational constant are 
one.

\section{Summary of Methods}
\label{method}

\subsection{Geometrical Variables}

In this paper, we solve time evolution 
based on the so called BSSN formalism~\cite{Shibata:1995we,Baumgarte:1998te} 
with a method of 2nd order finite differences. 
The maximal slicing and the hyperbolic gamma driver~\cite{Alcubierre:2002kk} 
are adopted for the gauge conditions. 
For stable calculations, we implement the Kreiss-Oliger~\cite{Kreiss-Oliger} 
dissipation. 
Although we do not write all equations down, 
just to fix the notation, 
we start with introducing geometrical and matter variables 
for the numerical integration. 
Readers may refer to several textbooks on numerical relativity(e.g., Refs.~\cite{2010nure.book.....B,Gourgoulhon:2007ue,shibata2016numerical}) for details.

We consider the following form of line elements:
\begin{equation}
\dd s^2=-\alpha^2 \dd t^2+\gamma_{ij} 
	\left(\dd x^i+\beta^i \dd t\right)
	\left(\dd x^j+\beta^j \dd t\right), 
\end{equation}
where $i,j=1,2,3$, and $\gamma_{ij}$, $\alpha$ and $\beta^i$ are 
the spatial metric, lapse function and shift vector, respectively. 
The Roman indices are lowered and raised by the spatial metric $\gamma_{ij}$.
For numerical integration, we use the Cartesian coordinate system 
and decompose the spatial metric as 
\begin{equation}
\gamma_{ij}=\ee^{4\psi}\tilde \gamma_{ij}~~{\rm with}~~\det \tilde \gamma=1. 
\end{equation}
The unit normal vector field $n^\mu$ to the spatial hyper-surface 
is given by $n_\mu=-\alpha\left(\dd t\right)_\mu$, 
where the Greek index $\mu$ runs from 0 to 3. 
Then, the projection tensor 
$\gamma_\mu^{~\nu}$ satisfying $\gamma_\mu^{~\nu}n_\nu=0$ is given by 
\begin{equation}
\gamma_\mu^{~\nu}=n_\mu n^\nu+g_\mu^{~\nu}. 
\end{equation}
The extrinsic curvature $K_{ij}$ is defined by 
\begin{equation}
K_{ij}=-\gamma_i^{~\mu}\gamma_j^{~\nu} \nabla_\mu n_\nu. 
\end{equation}
We adopt the following decomposition of the extrinsic curvature:
\begin{equation}
K_{ij}=\ee^{4\psi}\tilde A_{ij}+\frac{1}{3}K\gamma_{ij}, 
\end{equation}
where $K$ and $\ee^{4\psi}\tilde A_{ij}$ are 
the trace and traceless parts of the extrinsic curvature $K_{ij}$. 
Then, the Einstein equations can be written in terms of $\alpha$, $\beta^i$, 
$\gamma_{ij}$($\tilde \gamma_{ij}$ and $\psi$) and $K_{ij}$($\tilde A_{ij}$ and $K$). 
For example, the Hamiltonian and momentum constraints are written as
\begin{eqnarray}
16\pi E-R-K^2+K_{ij}K^{ij}&=&0,
\label{eq:Hamcon0}
\\
\mathcal M^i:=8\pi J^i-D_jK^{ji}+D^iK&=&0, 
\label{eq:Momcon0}
\end{eqnarray}
where $R$ and $D_i$ are the scalar curvature and the covariant derivative 
with respect to $\gamma_{ij}$, and 
$E$ and $J^i$ are defined by using the stress energy tensor $T^{\mu\nu}$ as follows: 
\begin{eqnarray}
E&=&n_\mu n_\nu T^{\mu\nu}, 
\label{eq:rhon}
\\
J^i&=&-\gamma^i_{~\mu} n_\nu T^{\mu\nu}. 
\label{eq:ji}
\end{eqnarray}
For a later convenience, we also introduce the following variable:
\begin{equation}
S^{ij}=\gamma^i_{~\mu}\gamma^j_{~\nu}T^{\mu\nu}. 
\label{eq:sij}
\end{equation}

\subsection{Stress energy tensor for a collisionless particle system}


Let us consider the collisionless particle system composed of 
$N$ particles each of which travels a timelike geodesic. 
The four velocity $u^\mu_p$ of the particle labelled by a positive integer $p$ 
can be decomposed as follows~\cite{Vincent:2012kn}:
\begin{equation}
u_p^\mu=\Gamma_p\left(n^\mu+V_p^\mu\right), 
\end{equation}
where the spatial velocity components $V_p^\mu$ satisfy 
$V_p^\mu n_\mu=0$%
, and $\Gamma_p$ is the Lorentz factor. 
Then, the 3+1 decomposition of the geodesic equations is 
expressed as follows~\cite{Vincent:2012kn}:
\begin{eqnarray}
\frac{\dd \Gamma_p}{\dd t}&=&\Gamma_p V_p^i\left(\alpha K_{ij} V_p^j-\del_i \alpha\right), \\
\frac{\dd V_p^j}{\dd t}&=&\alpha V_p^j 
\left[V_p^i \left(\del_j\ln \alpha-K_{jk}V_p^k\right)
+2K^i_{~j}-V_p^k\Gamma^i_{jk}\right]
-\gamma^{ij}\del_j\alpha-V_p^j\del_j\beta^i,\\
\frac{\dd \tau_p}{\dd t}&=&\frac{\alpha}{\Gamma_p},
\end{eqnarray}
where $\tau_p$ is the proper time and $\dd/\dd t=\frac{\alpha}{\Gamma_p}u_p^\mu \del_\mu$. 

The energy momentum tensor for a particle system 
is given by (see, e.g., \cite{Misner:1974qy})
\begin{equation}
T^{\mu\nu}=-\sum_p m_p
\frac{\delta^3\left(\bm x-\bm x_p\right)}{u^\lambda_p n_\lambda \sqrt{\gamma}}
u^\mu_pu^\nu_p, 
\end{equation}
where $m_p$ is the proper mass of the particle and $\bm x$ and $\bm x_p$ denote 
the spatial coordinates and those values at the particle position, respectively. 
Then, from Eqs.~(\ref{eq:rhon}--\ref{eq:sij}), we obtain  
\begin{eqnarray}
E&=&
\sum_p m_p \Gamma_p\frac{\delta^3\left(\bm x -\bm x_p\right)}{\sqrt{\gamma}},
\label{eq:rhon2}
\\
J^i&=&\sum_p m_p \Gamma_p V^i_p \frac{\delta^3\left(\bm x-\bm x_p\right)}{\sqrt{\gamma}},
\label{eq:ji2}
\\
S^{ij}&=&\sum_pm_p \Gamma_p V^i_p V^j_p 
\frac{\delta^3\left(\bm x-\bm x_p\right)}{\sqrt{\gamma}}. 
\label{eq:sij2}
\end{eqnarray}
In this paper, we assume that the proper mass of every particle is identical to $m$. 

Since the delta function cannot be numerically treated, 
we introduce the following smoothing: 
\begin{equation}
\delta^3\left(\bm x-\bm x_p\right)
\rightarrow
f_{\rm sp}\left(\left|\bm x-\bm x_p\right|, r_{\rm s}\right)
\end{equation}
with   
\begin{equation}
f_{\rm sp}(r,r_{\rm s})=\frac{1}{\pi r_{\rm s}^3}
\left\{
\begin{array}{lll}
1-\frac{3}{2}\left(\frac{r}{r_{\rm s}}\right)^2
+\frac{3}{4}\left(\frac{r}{r_{\rm s}}\right)^3&{\rm for}
&0\leq \frac{r}{r_{\rm s}} \leq 1
\\
\frac{1}{4}\left(2-\frac{r}{r_{\rm s}}\right)^3&{\rm for}
&1 \leq \frac{r}{r_{\rm s}} \leq 2
\\
0&{\rm for}&2<\frac{r}{r_{\rm s}}
\end{array}\right., 
\end{equation}
where $r_{\rm s}$ characterizes the size of a particle. 

\subsection{Cleaning of the Hamiltonian constraint violation}

In order to reduce the violation of the Hamiltonian constraint, 
we update the conformal factor $\psi$ at each time step. 
The update is done by using the iteration steps of 
the Successive Over-Relaxation (SOR) method for solving 
the elliptic equation of the Hamiltonian constraint. 
The Hamiltonian constraint \eqref{eq:Hamcon0} can be rewritten in 
the following form:
\begin{equation}
\tilde D_i{\tilde D}^i\psi
=-\tilde D_i\psi {\tilde D}^i\psi+\frac{1}{8}\tilde R +\ee^{4\psi}
\left(\frac{1}{12}K^2-\frac{1}{8}\tilde A_{ij}{\tilde A}^{ij}
-2\pi E\right),  
\label{eq:clean}
\end{equation}
where $\tilde D_i$ and $\tilde R$ are the covariant derivative and Ricci scalar 
with respect to $\tilde \gamma_{ij}$. 
The iteration step is repeated only a few times 
depending on the degree of the violation. 
This prescription reduces the violation of the Hamiltonian constraint during 
the time evolution. 

\subsection{Flow of time evolution}

We use the 2nd order leap frog method with time filtering for the time evolution. 
In our calculation, we slightly modified 
the evolution of ${\tilde \Gamma}^i$ defined by $-\del_i {\tilde \gamma}^{ij}$ 
from that in the conventional BSSN scheme as follows:
\begin{equation}
\frac{\del{\tilde \Gamma}^i}{\del t}=\left[\frac{\del{\tilde \Gamma}^i}{\del t}\right]_{\rm BSSN}
-2\ee^{4\psi}\mathcal M^i, 
\end{equation}
where the first term in the right-hand side represents the conventional 
terms in the BSSN scheme%
(see, e.g., Refs.~\cite{2010nure.book.....B,Gourgoulhon:2007ue,shibata2016numerical}). 
This modification does not make any qualitative difference in the results, 
but reduces the momentum constraint violation by a factor of a few in our simulation(see, e.g., Refs.~\cite{Frittelli:1996wr,Kidder:2001tz} for similar prescriptions). 
It should be noted that the added term is trivial if the momentum constraints are 
well satisfied. 
The reason for the smaller momentum constraint violation is not clear 
and further careful investigation would be needed 
for other practical application of this procedure. 
But, we do not pursue the reason further in this paper
since the modification does not make any 
qualitative difference in our case.

The flow of the calculation is as follows:
\begin{enumerate}
\item{Starting from the initial data, we calculate 
next step geometrical variables except for $\alpha$. 
Variables for each particle are also evolved by using 
the geodesic equations, where geometrical variables 
at each particle position are calculated by 
using a 2nd order interpolation. }

\item{The energy momentum tensor is calculated from the particle distribution. 
Here, we note that the expressions (\ref{eq:rhon}--\ref{eq:sij}) 
are independent of $\alpha$ which has not been fixed yet. }

\item{$\psi$ is updated by the cleaning of the Hamiltonian constraint violation. }

\item{By solving the elliptic equation of the maximal slice condition, 
we obtain $\alpha$. }
\end{enumerate}
The above procedure is repeated.

\section{Shapiro-Teukolsky Collapse with particles}

\subsection{Initial Data Construction}
As in Ref.~\cite{Shapiro:1991zza}, we start with conformally flat 
and momentarily static initial data, that is, 
\begin{equation}
\tilde \gamma_{ij}=\delta_{ij}~,~~K_{ij}=0.  
\end{equation}
The momentum constraint equation \eqref{eq:Momcon0} is trivially satisfied by setting $J^i=0$. 
In terms of particle variables, we assume 
$\Gamma_p=1$ and $V^i_p=0$ for every particle. 
The Hamiltonian constraint equation is written as 
\begin{equation}
\triangle \Psi = -2\pi E \Psi^5
=-2m \sum_p f_{\rm sp}\left(\left|\bm x-\bm x_p\right|, r_{\rm s}\right)/\Psi,
\label{eq:hamcon}
\end{equation}
where $\Psi=\ee^\psi$ and $\triangle$ is the Laplace operator in 
the 3-dimensional Euclidean space. 
This equation can be solved by using SOR method 
once the particle distribution is fixed. 
We generate the particle distribution 
with reference to a continuous density distribution and 
the corresponding conformal factor 
denoted by $\bar E$ and $\bar \Psi$, respectively.

Following Refs.~\cite{Shapiro:1991zza,Nakamura:1988zq}, 
we determine the particle distribution 
based on the following continuum density distribution:
\begin{equation}
\frac{1}{2}\bar E\bar \Psi^5=
E_{\rm N}:=\left\{
\begin{array}{lll}
\frac{3M_{\rm N}}{4\pi a^2 b}&{\rm for}
&\frac{x^2+y^2}{a^2}+\frac{z^2}{b^2}\leq 1
\\
0&{\rm for}
&\frac{x^2+y^2}{a^2}+\frac{z^2}{b^2}> 1
\end{array}\right. , 
\end{equation}
where $M_{\rm N}$ is a constant which represents the total 
Newtonian rest mass, $a$ is the equatorial radius and 
$b$ is the radius of the major axis. 
Then, $\bar \Psi$ can be expressed by the Newtonian potential $\Phi$ 
as follows:
\begin{equation}
\bar \Psi=1-\Phi 
\end{equation}
with
\begin{equation}
\triangle \Phi = 4\pi E_{\rm N}. 
\end{equation}
For a prolate ($a<b$) spheroid, 
we obtain~\cite{1969efe..book.....C,Nakamura:1988zq} 
\begin{eqnarray}
\Phi&=&
-\frac{3M_{\rm N}}{2be}\beta
-\frac{3M_{\rm N}}{4b^3e^3}\left(\beta-\sinh\beta\cosh\beta\right)R^2
-\frac{3M_{\rm N}}{2b^3e^3}\left(\tanh\beta-\beta\right)z^2, 
\end{eqnarray}
where $e=\sqrt{1-a^2/b^2}$ and $R=\sqrt{x^2+y^2}$, 
and $\beta$ satisfies 
\begin{equation}
\begin{array}{lll}
\sinh\beta=\frac{be}{a}
&{\rm for~}&\frac{x^2+y^2}{a^2}+\frac{z^2}{b^2}\leq 1, \\
R^2\sinh^2\beta+z^2\tanh^2\beta=b^2e^2
&{\rm for~}&\frac{x^2+y^2}{a^2}+\frac{z^2}{b^2}> 1.  
\end{array}
\end{equation}
%
%
%
%
Since the asymptotic behaviour in the limit $r=\sqrt{x^2+y^2+z^2}\rightarrow \infty$ 
is given by 
\begin{equation}
\lim_{r\rightarrow \infty}\bar \Psi
=1-\lim_{r\rightarrow \infty}\Phi=1+\frac{M_{\rm N}}{r}, 
\end{equation}
taking the isotropic coordinate into account, 
the total mass $M$ can be read off as $2M_{\rm N}$.

By using this density distribution as the reference, 
the number of particles $\Delta N$ in a grid box with 
the volume $\Delta V$ is set by 
\begin{equation}
\Delta N
=\frac{\bar E \bar \Psi^6}{m}\Delta V 
=\frac{2E_{\rm N}\bar \Psi}{m}\Delta V, 
\label{eq:dN}
\end{equation}
where the mass $m$ is related to the total number of particles $N$ as 
\begin{equation}
mN=M_0:=\int \bar E \bar \Psi^6 \dd^3 x
=\int 2E_{\rm N}\bar \Psi \dd^3 x
=2M_{\rm N}+\frac{6}{5}\frac{M_{\rm N}^2}{be}\ln\frac{1+e}{1-e}. 
\end{equation}
We randomly distribute $N$ particles in accordance with Eq.~\eqref{eq:dN}, 
and numerically solve Eq.~\eqref{eq:hamcon}. 
Here, we note that, whereas the reference density distribution of the continuum
is the same as that in ST, 
exact axisymmetry is not assumed 
in our case unlike in the ST case. 
This is because the real density distribution is composed of 
the particles which are randomly distributed.

\subsection{Results for the same parameter setting as ST}
\label{subsec:ST}
We consider the domain for the numerical calculation given by 
$0<x^i/L<1$, where $x^i=(x,y,z)$. 
Hereafter, we normalize all dimensionful quantities in the unit of $M$. 
We consider the situation characterized by the following parameter set:
\begin{equation}
\frac{L}{M}=20~,~~
\frac{b}{M}=10~,~~
e=0.9. 
\end{equation}
The values of $b/M$ and $e$ are equivalent to those in ST.  
The initial data given by this parameter set result in 
a spindle collapse without an apparent horizon. 
In the calculation, the grid interval $\Delta$, 
particle size $r_{\rm s}$ and particle number $N$ are set as 
\begin{equation}
\Delta=L/120~,~~N=10^6~,~~ r_{\rm s}=L/75.
\end{equation}
We have also performed a set of simulations of the physically identical
model with several different resolutions. 
Changing the grid interval $\Delta$, we impose the following scaling for 
the particle size $r_{\rm s}$ and number $N$: 
\begin{equation}
r_{\rm s}\propto \Delta~,~~N\propto 1/\Delta^3,
\end{equation}
so that $\Delta N$, the number of particles in a grid box, 
is kept constant.  
We have checked that the dynamics of the system does not significantly depend 
on the size and shape of the particle profile(see Ref.~\cite{Yamada:2011br} 
for results with a Gaussian shape).
All figures are for the case of $\Delta=L/120$ unless otherwise noted. 
We note that, in order to minimize the dispersion in dependence on the resolution, 
we use a common pseudo random numbers 
to generate particle distribution for each resolution. 
Therefore, a part of particles have identical initial positions for each resolution. 

\begin{figure}[htbp]
\begin{center}
\includegraphics[scale=1]{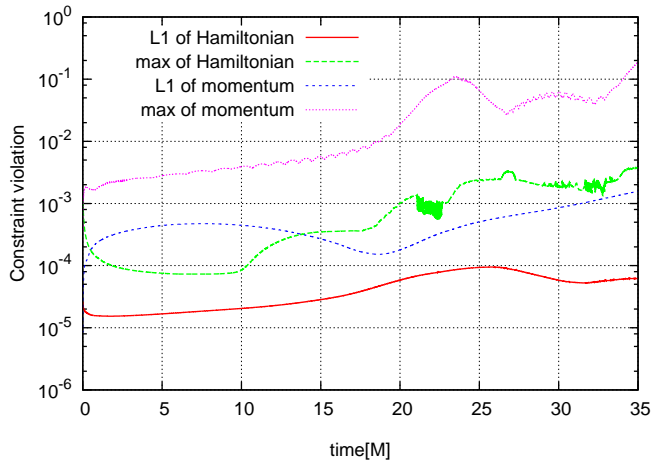}
\caption{Constraint violations are depicted as functions of time. The
violation of the Hamiltonian constraint and the momentum constraints 
is evaluated at each grid point using an appropriate normalization. 
The L1-norm and max-norm are calculated for each constraint and 
plotted as functions of time. 
}
\label{fig:consts}
\end{center}
\end{figure}

\begin{figure}[htbp]
\begin{center}
\includegraphics[scale=1]{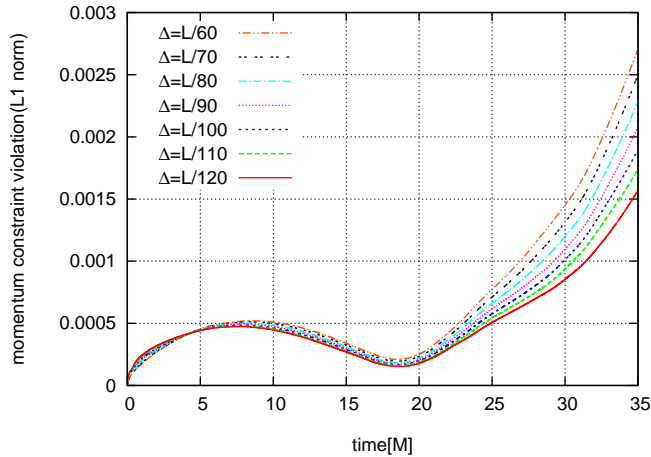}
\caption{L1-norm of the momentum constraint violation 
is depicted as a function of time for each 
resolution. 
}
\label{fig:convconst}
\end{center}
\end{figure}

We emphasize that we have monitored the existence of an apparent horizon
covering the origin of the coordinates during the time evolution starting from 
the initial data given in the previous section 
and concluded that there is no horizon during the time evolution. 
On the other hand, as will be shown in the next subsection, 
starting from a different initial data set, 
we have found an example in which an apparent horizon is finally formed 
after a spindle collapse. 
We have also monitored the violation of constraint equations (Fig.~\ref{fig:consts}). 
%
We found that suppressing the max-norm of the momentum constraint violation 
is relatively hard with our numerical scheme. 
In this subsection, we require that the max-norm of the momentum constraint violation 
is at most at the level of 10\%. 
The resolution dependence of the L1-norm of the momentum constraint violation 
is shown in Fig.~\ref{fig:convconst}. 
It should be noted that, since we scale the total number and 
size of particles depending on the grid interval, 
the usual second order convergence cannot be expected(see Appendix for 
the second order convergence with fixed 
number and size of particles). 
Nevertheless, Fig.~\ref{fig:convconst} shows that 
the constraint violation is smaller for a finer resolution
at late times. 
We also note that in our simulation the convergence is not 
clear for local quantities without averaging 
due to the randomness of the
particle distribution.

First, we show the snapshots of the particle distribution and density distribution, 
the values of the Kretschmann curvature invariant 
$\mathcal K:=R_{\mu\nu\rho\lambda}R^{\mu\nu\rho\lambda}$ and 
the Weyl curvature invariant $\mathcal W:=C_{\mu\nu\rho\lambda}C^{\mu\nu\rho\lambda}$ 
on the $x$-$z$ plane, where $C_{\mu\nu\rho\lambda}$ is the Weyl curvature. 
\begin{figure}[htbp]
\begin{center}
\includegraphics[scale=0.7]{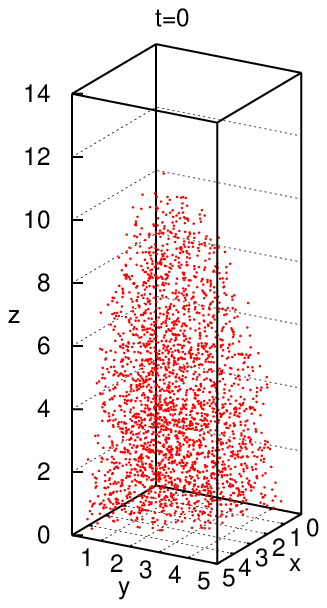}~
\includegraphics[scale=0.7]{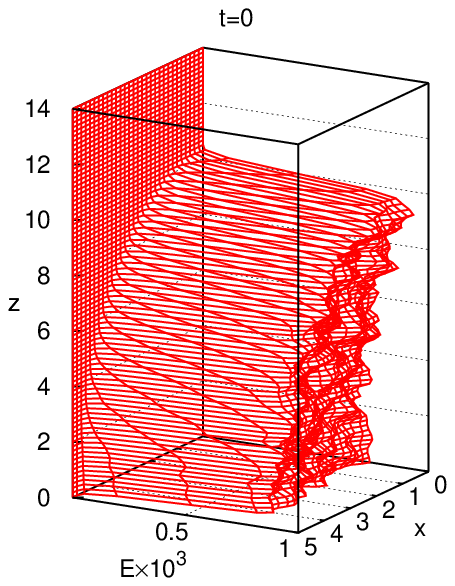}~
\includegraphics[scale=0.7]{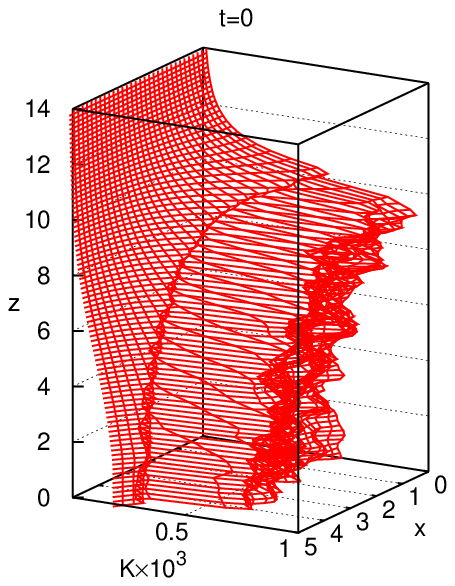}~
\includegraphics[scale=0.7]{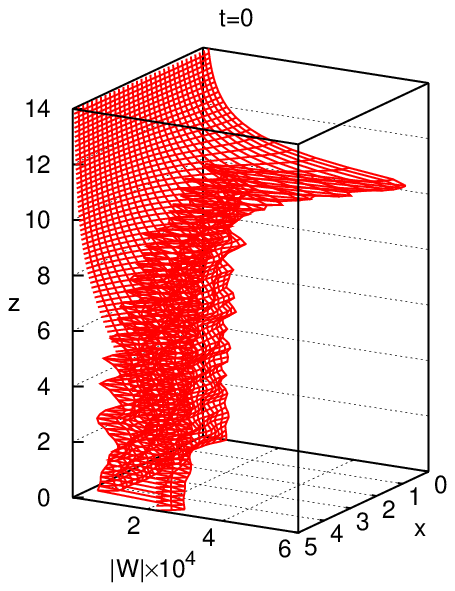}
\vspace{3mm}

\includegraphics[scale=0.7]{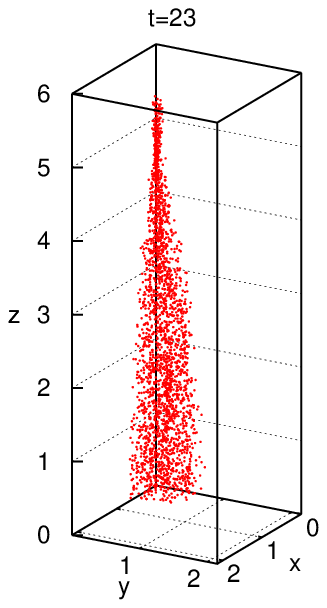}~
\includegraphics[scale=0.7]{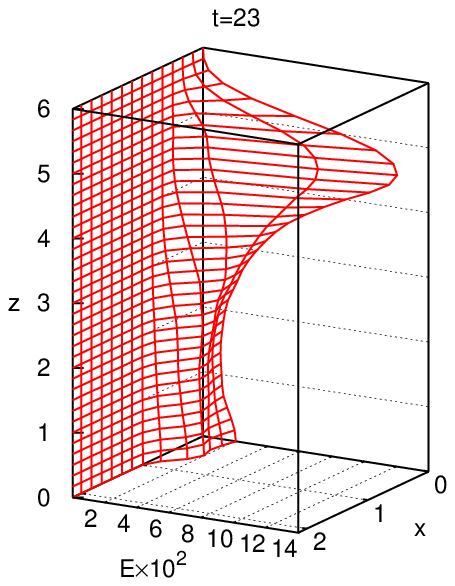}~
\includegraphics[scale=0.7]{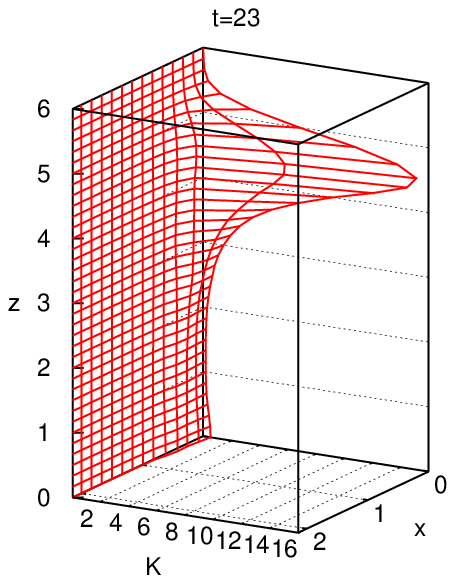}~
\includegraphics[scale=0.7]{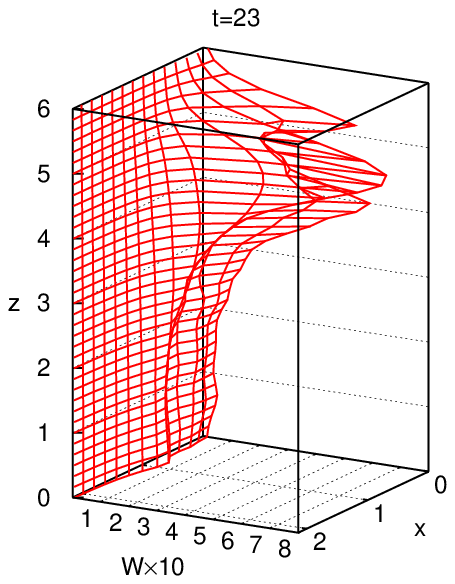}
\vspace{3mm}

\includegraphics[scale=0.7]{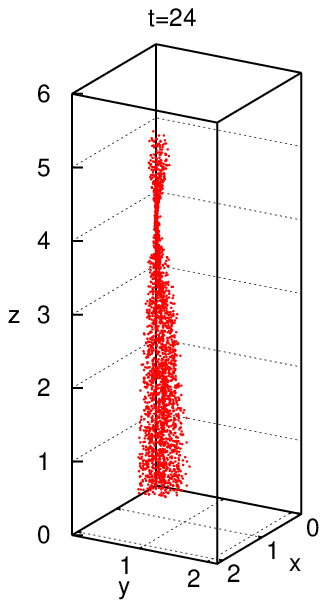}~
\includegraphics[scale=0.7]{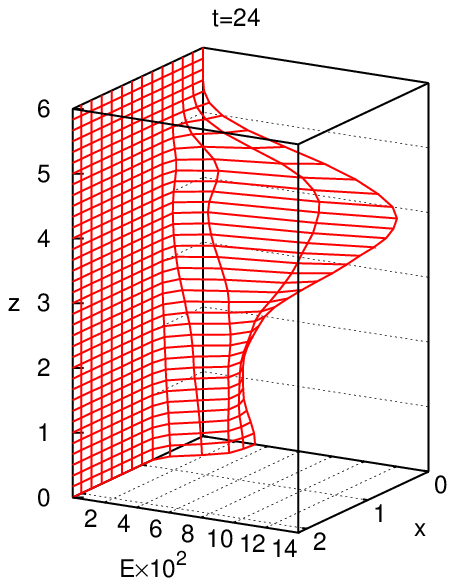}~
\includegraphics[scale=0.7]{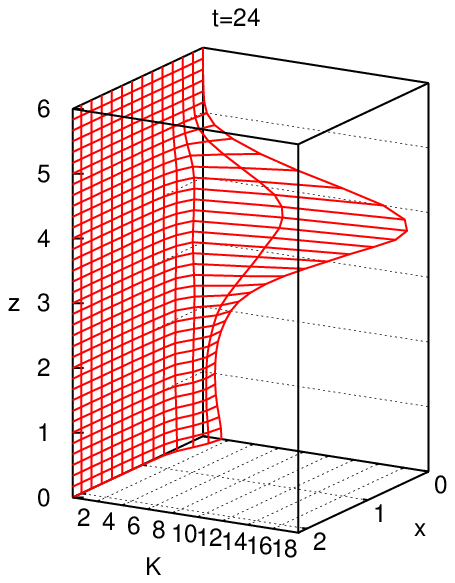}~
\includegraphics[scale=0.7]{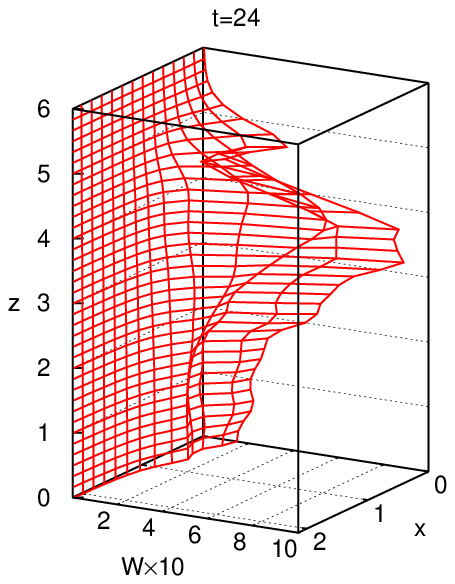}
\vspace{3mm}

\includegraphics[scale=0.7]{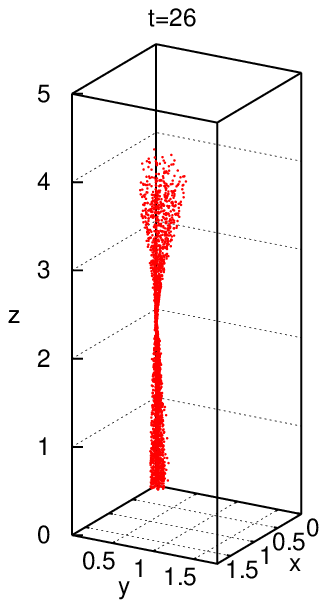}~
\includegraphics[scale=0.7]{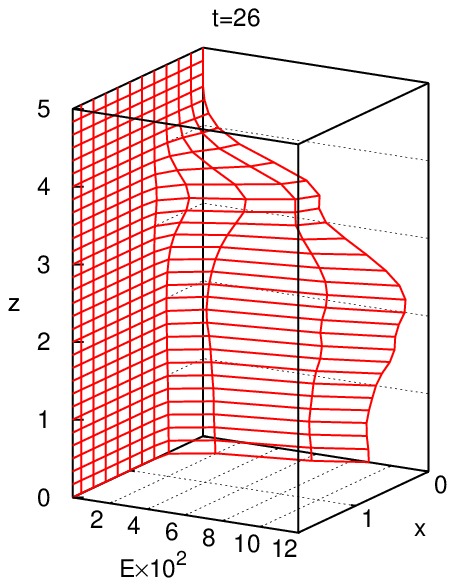}~
\includegraphics[scale=0.7]{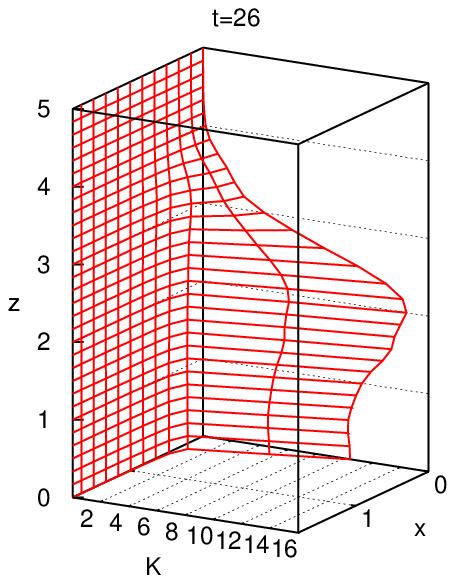}~
\includegraphics[scale=0.7]{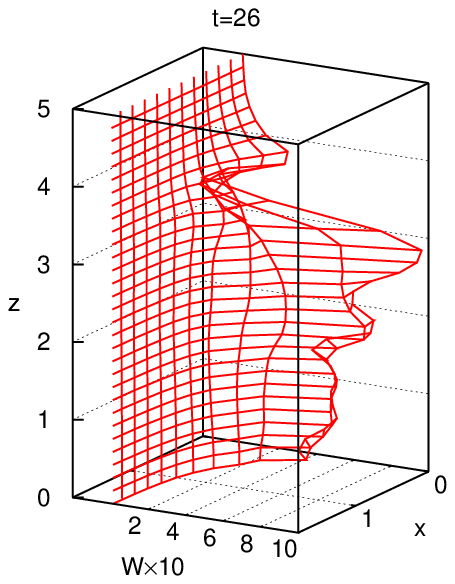}
\caption{
Snapshots of the particle distribution and density distribution, 
Kretschmann curvature invariant $\mathcal K$ and 
Weyl curvature invariant $\mathcal W$ on the $x$-$z$ plane 
are shown. 
The numerical value of every dimensionful variable 
is given in the unit of $M$. 
The snapshots at $t=23M$ are shown for comparison with figures in ST. 
The peak values $\mathcal K_{\rm p}$ and $\mathcal W_{\rm p}$ take 
the maximum values at $t\simeq 24M$ and $t\simeq 25M$, respectively. 
}
\label{fig:evolution}
\end{center}
\end{figure}
%
Since our shift gauge condition is different from 
that used in ST, the spatial shape of the particle distribution 
in our coordinates cannot be directly compared with that in ST 
in the strict sense. 
Nevertheless, as 
is shown in Fig.~ \ref{fig:evolution}, at $t=23M$, 
we can find a matter concentration near $z=5M$
similarly to the result in ST. 
Around this time, the system experiences the first caustic 
near the top of the matter distribution. 
We note that, unlike in ST~\cite{Shapiro:1991zza}, 
our calculation does not break down even after this time. 
After the first caustic, the particles which went through the caustic 
start to spread outward from the $z$-axis. 
The total length of $z$-direction continues to shrink, 
and the position of the caustic moves inward toward the origin.

\begin{figure}[htbp]
\begin{center}
\includegraphics[scale=0.9]{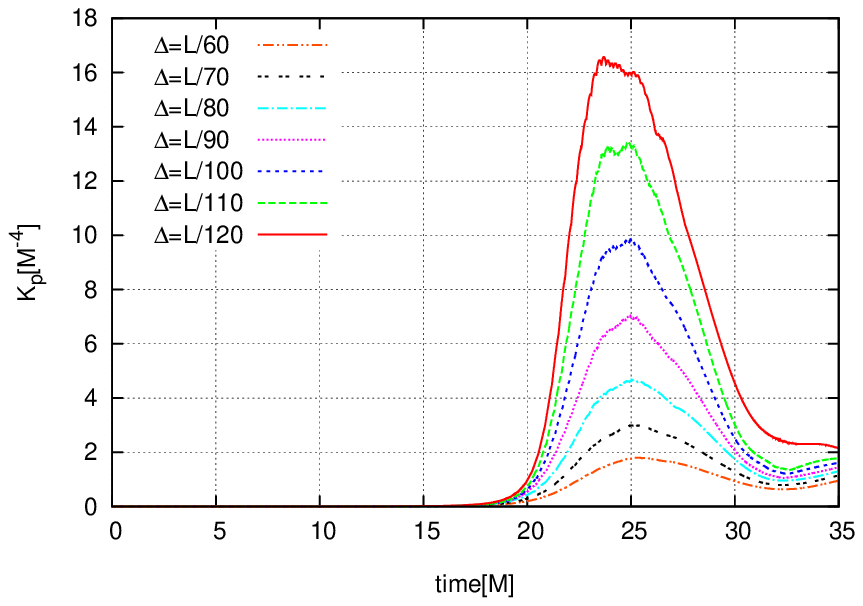}
\includegraphics[scale=0.9]{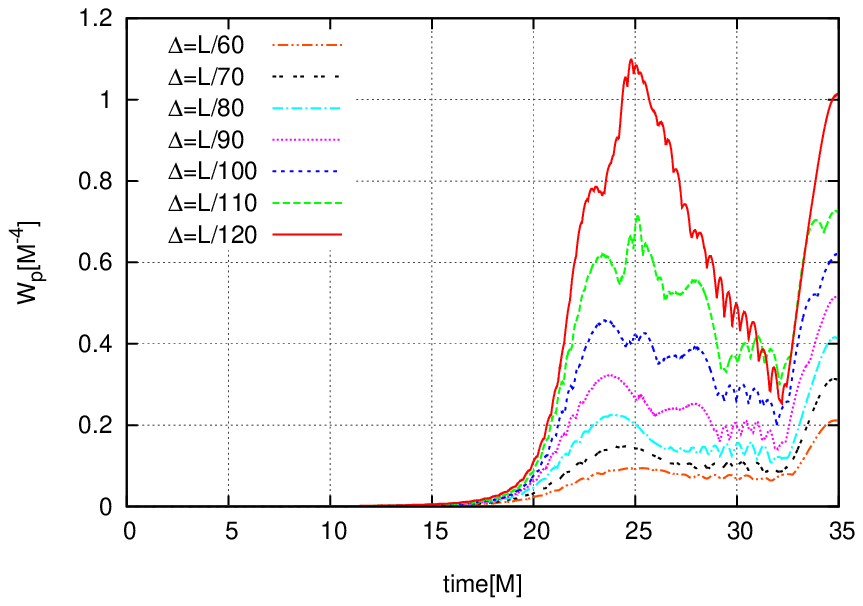}
\caption{$\mathcal K_{\rm p}$ and $\mathcal W_{\rm p}$ are depicted as functions of time. 
}
\label{fig:K}
\end{center}
\end{figure}

At $t= 24M$, 
the value of $\mathcal K$ has a peak at a point near the density peak. 
Around this time, the peak value $\mathcal K_{\rm p}$ 
takes the maximum value $\mathcal K_{\rm max}$ and 
gradually decreases with time after that. 
The value of $\mathcal W$ at its peak, denoted by $\mathcal W_{\rm p}$, 
takes the maximum value $\mathcal W_{\rm max}$ 
around $t=25M$ soon after the time when $\mathcal K_{\rm p}=\mathcal K_{\rm max}$. 
The values of $\mathcal K_{\rm p}$ and $\mathcal W_{\rm p}$ are 
depicted as functions of time in Fig.~\ref{fig:K}. 
As is similar to FIG.~3 in ST, the value of $\mathcal K_{\rm p}$ 
starts to increase around $t\simeq 20M$, 
and we find faster growth for a finer resolution. 
%

\begin{figure}[htbp]
\begin{center}
\includegraphics[scale=0.9]{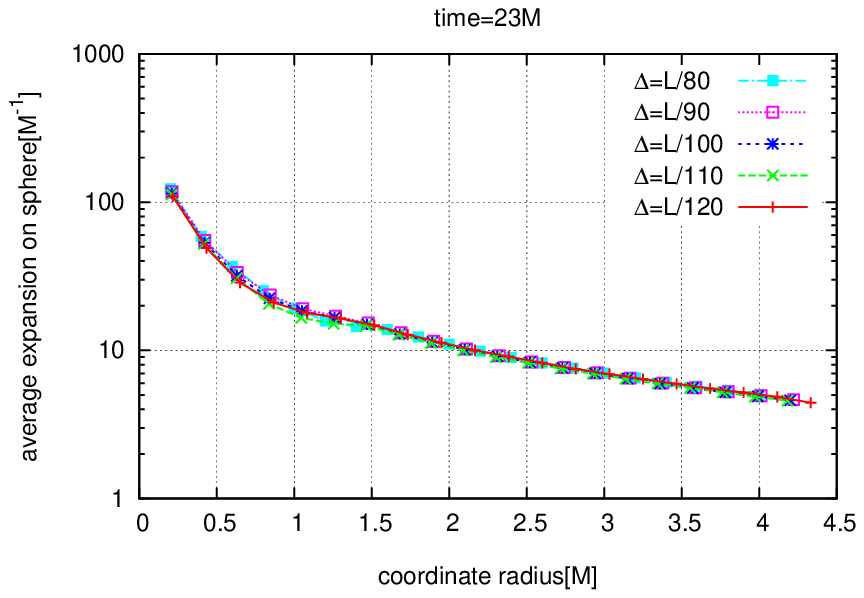}
\includegraphics[scale=0.9]{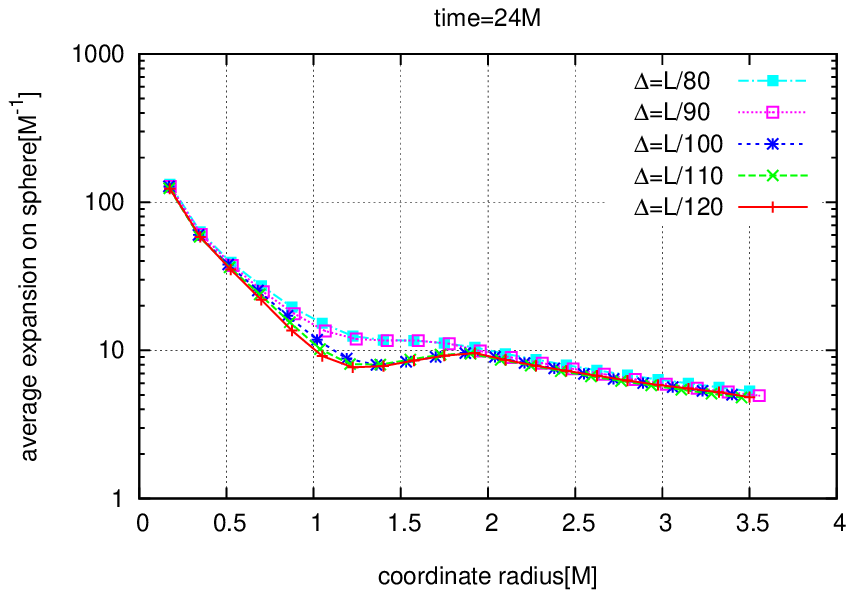}
\caption{The averaged value of $\Theta$ is depicted as a function 
of the radius of the sphere for each resolution. 
}
\label{fig:expansion}
\end{center}
\end{figure}

In spite of the dynamics qualitatively similar to that reported in ST, 
we can also find a significant difference from ST. 
In ST, it is reported that the peak position of the curvature invariant 
at $t=23M$ is at $\simeq 6.1M$ outside the matter distribution. 
While in our case, the functional form of $\mathcal K$ roughly traces 
the form of the density distribution. 
The main contribution for $\mathcal K_{\rm p}$ comes from the Ricci part of the 
curvature. 
Since the density is also divergent near the peak of $\mathcal K$, 
one may be concerned with a small trapped region in the vicinity 
of the peak. 
If the size of the trapped region is as small as a few grid intervals, 
our apparent horizon finder can not resolve it. 
Therefore, 
in order to investigate the existence of the small trapped region, 
we calculate the value of expansion $\Theta$ on the spheres 
centered at the peak of $\mathcal K$
instead of searching for the apparent horizon.
The expansion $\Theta$ is defined by 
\begin{equation}
\Theta=D_is^i+K_{ij}s^is^j-K, 
\end{equation}
where $s^i$ is the unit normal vector to the sphere. 
In Fig.~\ref{fig:expansion}, we depict the value of $\Theta$ 
averaged on each sphere as a function of the radius of the sphere. 
As is shown in Fig.~\ref{fig:expansion}, 
the average value of $\Theta$ is positive at least within our resolution. 
This suggests nonexistence of such an apparent horizon.

\begin{figure}[htbp]
\begin{center}
\includegraphics[scale=0.9]{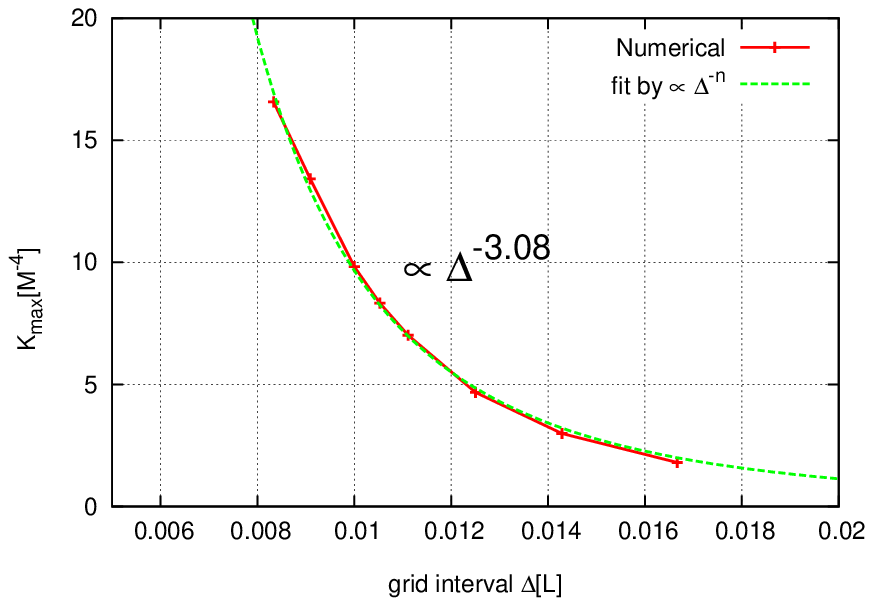}
\includegraphics[scale=0.9]{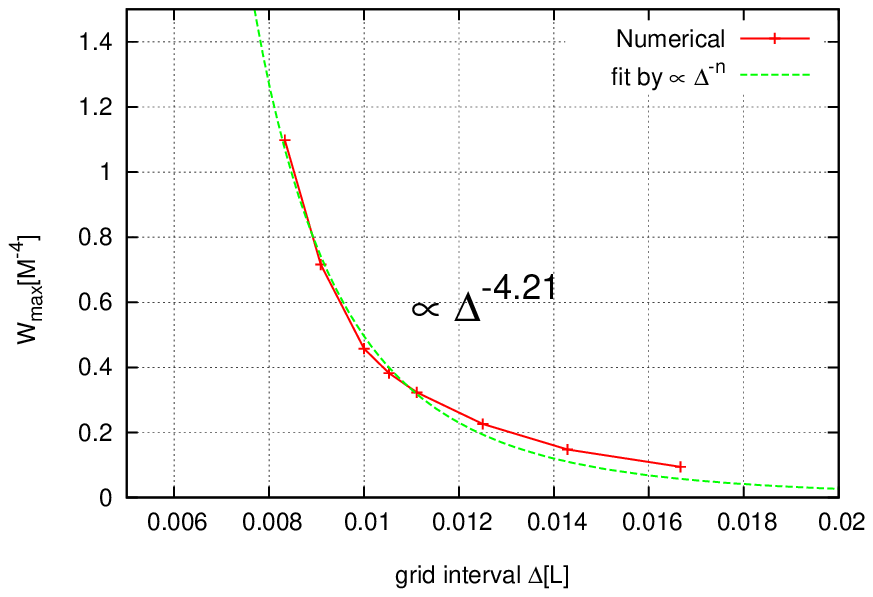}
\caption{$\mathcal K_{\rm max}$ and $\mathcal W_{\rm max}$ are depicted as functions of 
the grid interval $\Delta$. 
}
\label{fig:Kremax}
\end{center}
\end{figure}

Finally, let us check how $\mathcal K_{\rm max}$ and $\mathcal W_{\rm max}$ 
depend on the numerical resolution. 
In Fig.~\ref{fig:Kremax}, 
$\mathcal K_{\rm max}$ and $\mathcal W_{\rm max}$ are depicted as functions of 
the grid interval $\Delta$, 
where $\mathcal W_{\rm max}$ is evaluated within the time interval
$0\leq t<30M$. 
The behaviour of $\mathcal K_{\rm max}$ for smaller $\Delta$ seems to have  
an inverse power dependence on $\Delta$. 
We also find similar tendency for $\mathcal W_{\rm max}$. 
These dependences suggest divergent curvature invariants  
in the limit of infinite resolution similarly to ST. 

\subsection{Results for a larger mass system}
\label{subsec:horizon}
As is written in the section \ref{subsec:ST}, 
we did not find an apparent horizon for the 
same parameter setting as the ST case. 
In the sense of the hoop conjecture\cite{Thorne1972}, 
we expect that a trapped region can be more easily formed 
for a larger mass system with the same size and shape of the matter distribution. 
In this subsection, we show an example in which an
apparent horizon is formed after the occurrence of 
the maximum value of the Kretschmann invariant by 
increasing the total mass comparing the grid interval. 
It is also worthy to note that, since the total mass 
of the previous system is given by $M=6\Delta$, 
the corresponding horizon radius in the isotropic coordinate is 
given by $M/2=3\Delta$. 
Therefore, the resolution seems to be not enough to resolve 
the spherical horizon for the Schwarzschild black hole with the same mass $M$.  

Let us consider the particle distribution 
generated by the following parameter set:
\begin{equation}
\frac{L}{M}=\frac{13}{2}~,~~
\frac{b}{M}=\frac{13}{4}~,~~
e=0.9. 
\end{equation}
We leave the value of ellipticity unchanged but 
consider a smaller initial size compared with the total mass $M$. 
The expected typical horizon radius in the isotropic coordinates 
is given by $\frac{1}{2}M=\frac{1}{13}L=\frac{120}{13}\Delta$.

\begin{figure}[htbp]
\begin{center}
\includegraphics[scale=0.7]{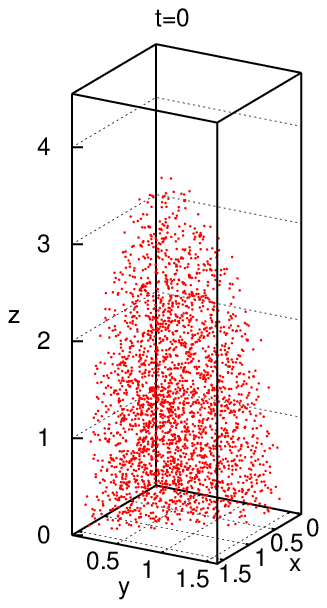}~
\includegraphics[scale=0.7]{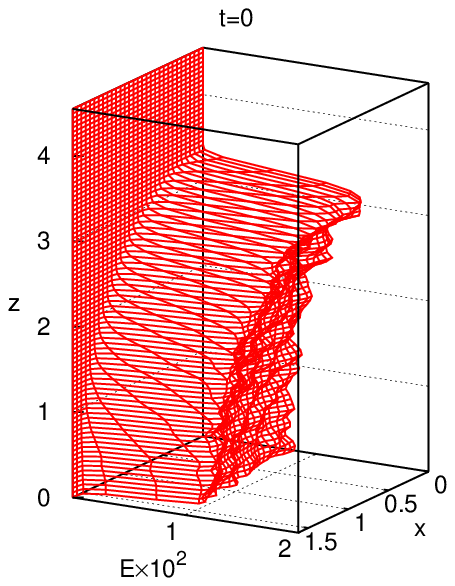}~
\includegraphics[scale=0.7]{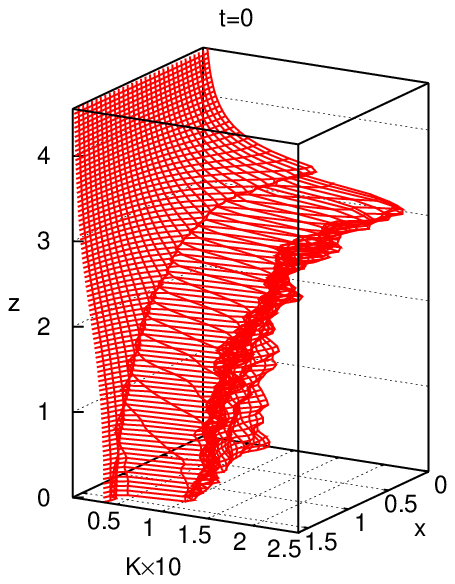}~
\includegraphics[scale=0.7]{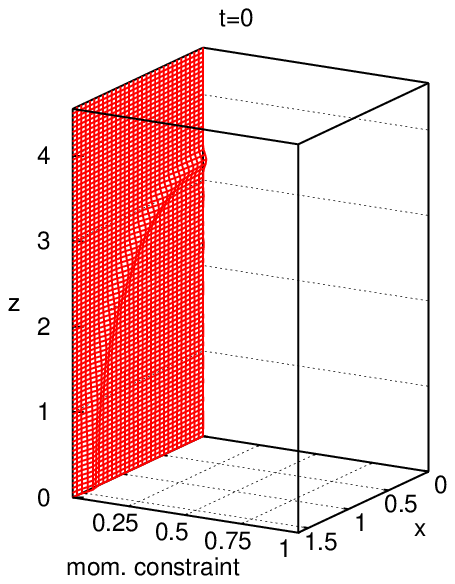}
\vspace{3mm}

\includegraphics[scale=0.7]{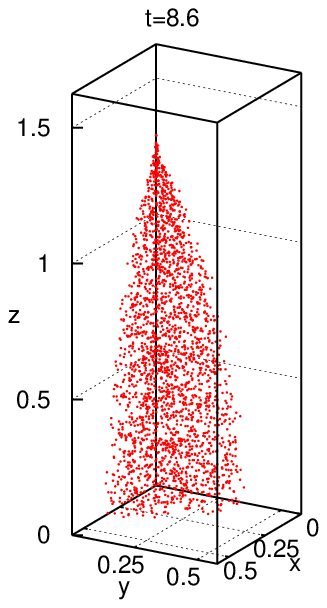}~
\includegraphics[scale=0.7]{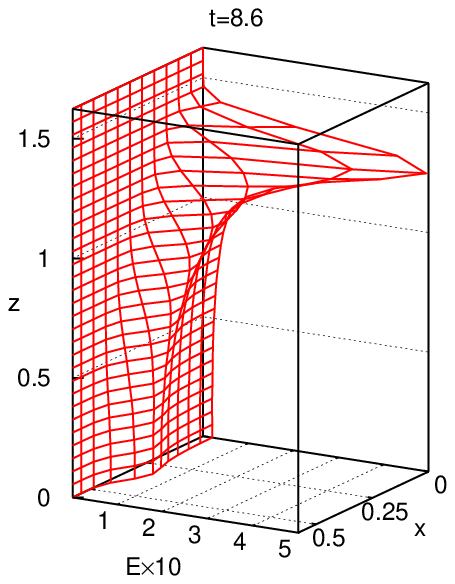}~
\includegraphics[scale=0.7]{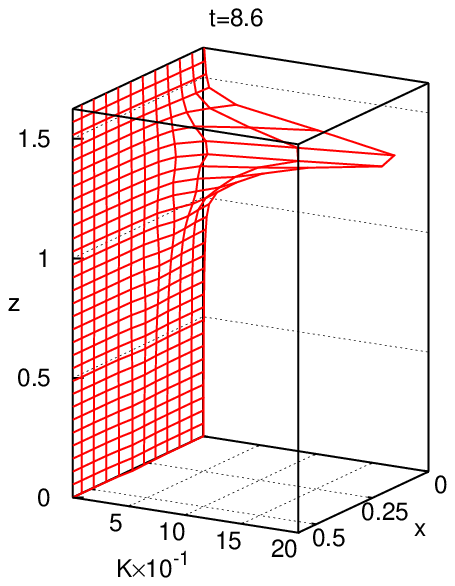}~
\includegraphics[scale=0.7]{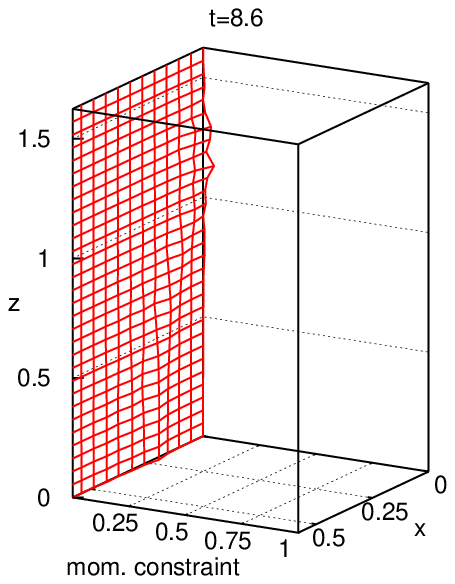}
\vspace{3mm}

\includegraphics[scale=0.7]{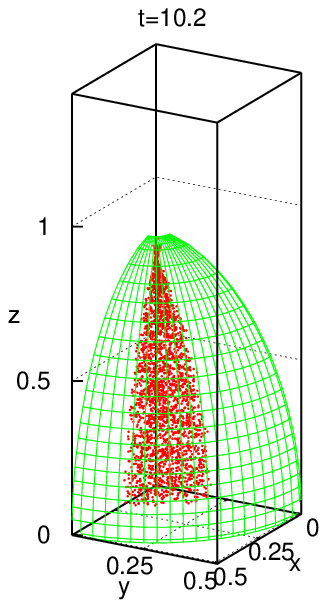}~
\includegraphics[scale=0.7]{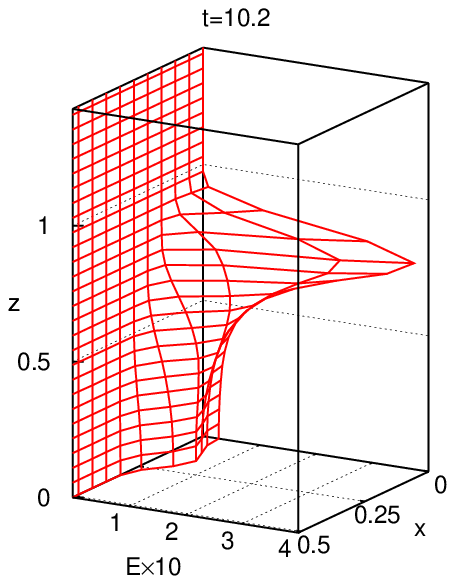}~
\includegraphics[scale=0.7]{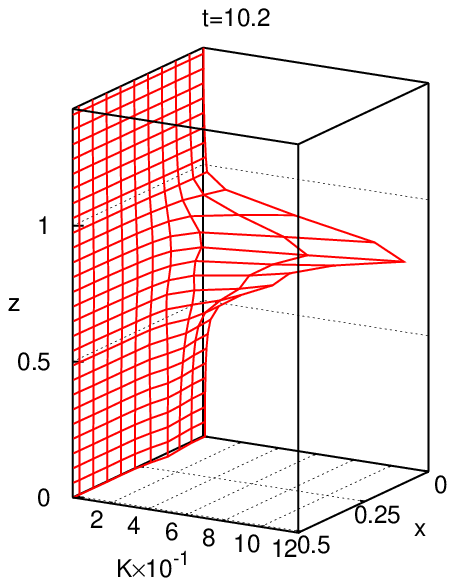}~
\includegraphics[scale=0.7]{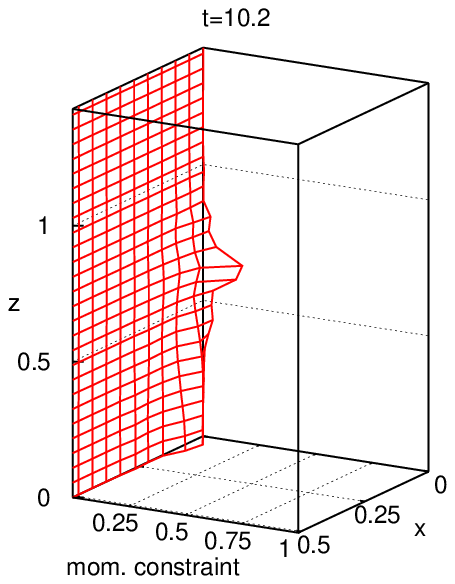}
\caption{
Snapshots of the particle distribution and the density distribution, 
Kretschmann curvature invariant $\mathcal K$ 
and momentum constraint violation on the $x$-$z$ plane 
are shown. 
The numerical value of every dimensionful variable 
is given in the unit of $M$. 
The peak value $\mathcal K_{\rm p}$ take 
the maximum values at $t\simeq 8.6M$. 
As is shown in the panel of the particle distribution
at $t=10.2M$, an apparent horizon appears at this time. 
}
\label{fig:evolution2}
\end{center}
\end{figure}

\begin{figure}[htbp]
\begin{center}
\includegraphics[scale=0.9]{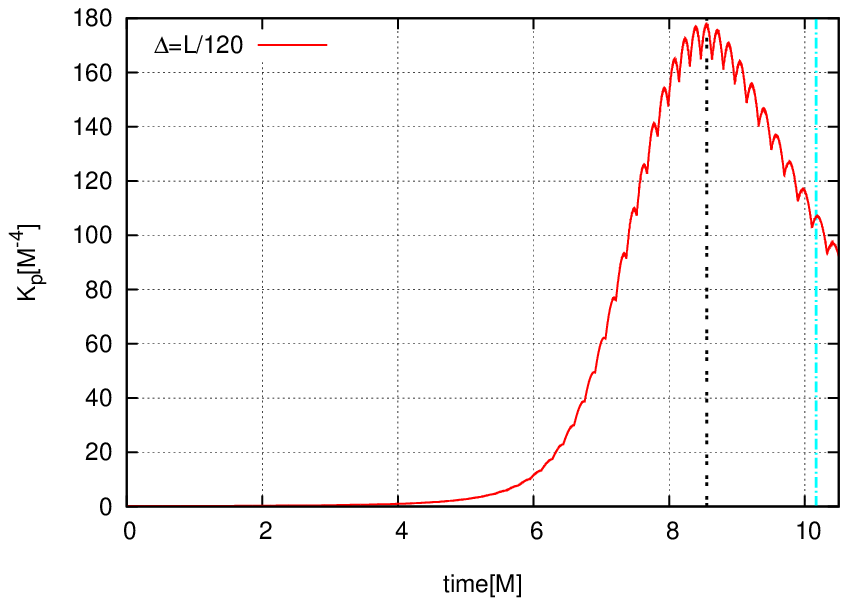}
\includegraphics[scale=0.9]{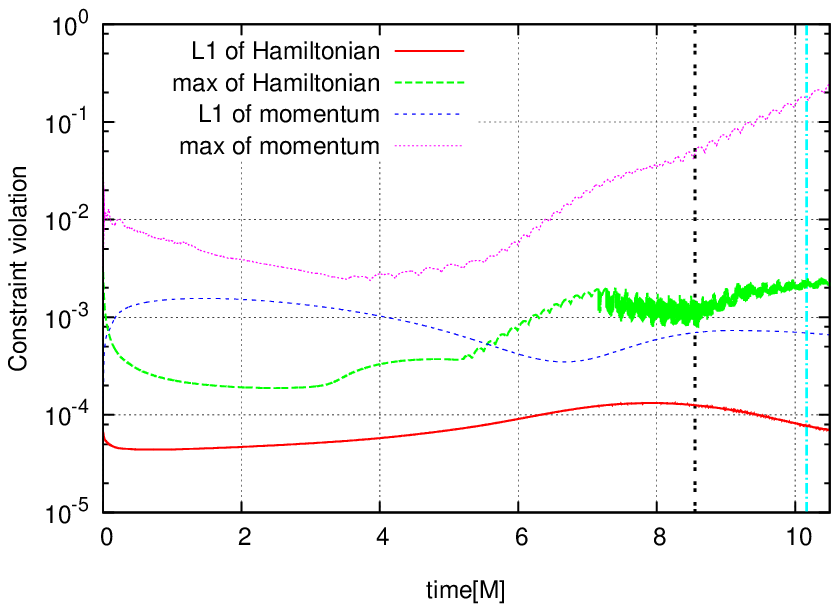}
\caption{$\mathcal K_{\rm p}$ and values of constraint violation 
are depicted as functions of time. 
The vertical lines show the time at $\mathcal K_{\rm max}$ and 
the horizon formation, respectively. 
}
\label{fig:Kpconst}
\end{center}
\end{figure}

\begin{figure}[htbp]
\begin{center}
\includegraphics[scale=1]{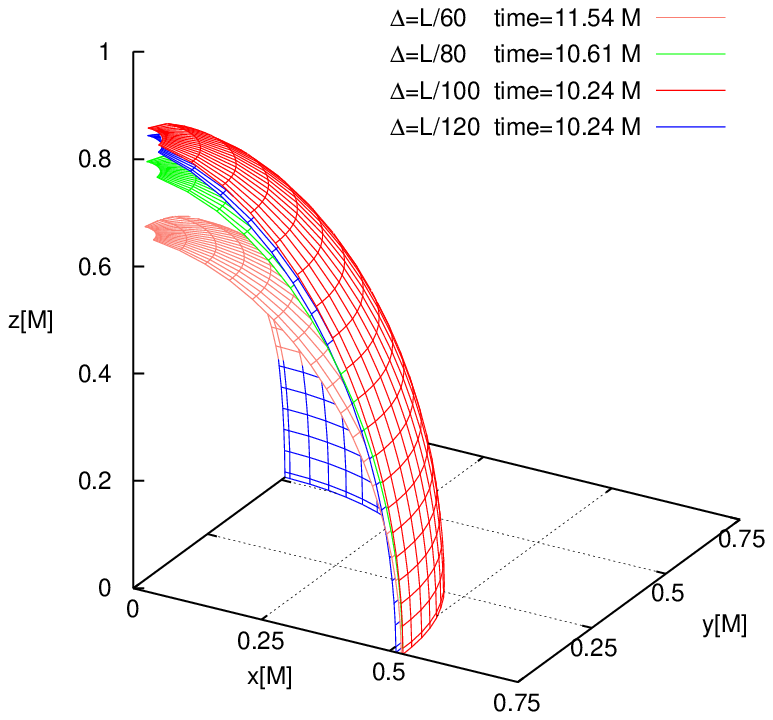}
\caption{Resolution dependence of the horizon formation time and 
the apparent shape are shown. 
}
\label{fig:horizons}
\end{center}
\end{figure}

\begin{figure}[htbp]
\begin{center}
\includegraphics[scale=1]{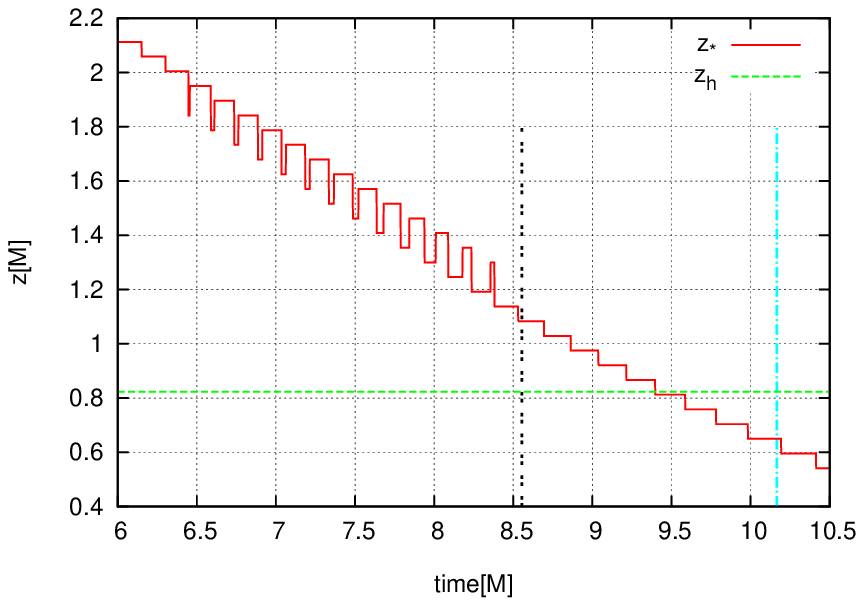}
\caption{$z_*$ is depicted as a function of time. 
The two vertical lines show the time at $\mathcal K_{\rm max}$ and 
the horizon formation, respectively. 
}
\label{fig:zstar}
\end{center}
\end{figure}

First, we show the snapshots of the particle distribution, 
density distribution, Kretschmann curvature invariant
and momentum constraint violation in Fig.~\ref{fig:evolution2}. 
The qualitative picture of the particle dynamics is similar to 
the previous case. 
However, in this case, an apparent horizon appears at $t\simeq 10.2M$ 
as is depicted in the lower left-most panel in Fig.~\ref{fig:evolution2}. 
The value of $z$-axis at the pole of the horizon, denoted by $z_{\rm h}$  
is given by $z_{\rm h}\simeq 0.82M$. 
The density, Kretschmann invariant and momentum constraint violation 
take the maximum value inside the horizon at the 
formation time. 
We show the time evolution of $\mathcal K_{\rm p}$ and constraint violation 
in Fig.~\ref{fig:Kpconst}. 
%
From this figure, it is clear that 
an apparent horizon is formed after $\mathcal K_{\rm p}$ takes the 
maximum value. 
We also checked the resolution dependence of 
the shape and time at the horizon formation time. 
As is shown in Fig.~\ref{fig:horizons}, 
the horizon formation time and the apparent shape are almost convergent.  
%
%
%
%
%
The position of the maximum momentum constraint violation is 
located on the $z$-axis after $t=6M$. 
Let $z_*$ denote the value of the $z$ coordinate at the peak 
of the momentum constraint violation.  
We depict the value of $z_*$ as a function of time in Fig.~\ref{fig:zstar}. 
Although $z_*$ is well inside the apparent horizon at the formation time, 
it takes a larger value at an earlier time. 

The result shown in this subsection is 
well understood in the sense of the hoop conjecture\cite{Thorne1972}. 
That is, the matter distribution is too elongated for 
the formation of the horizon which covers the most part of the system 
at the moment of $\mathcal K_p=\mathcal K_{\rm max}$.
However, after a certain period of time after $\mathcal K_p=\mathcal K_{\rm max}$, 
the matter distribution gets compacted into a region 
whose circumference in every direction is comparable to $4\pi M$. 
 

\section{Summary and Discussion}
We have performed 3D general relativistic simulations of the 
non-spherical gravitational collapse of a collisionless particle system. 
Particles have been randomly distributed inside a prolate spheroid.
Unlike the case done by Shapiro and Teukolsky~(ST) 
in Ref.~\cite{Shapiro:1991zza}, 
exact axisymmetry has not been assumed. 
We have found that 
a peak of the Kretschmann curvature invariant
appears near the pole of the matter distribution, 
and the peak value takes a maximum after a period of time. 
The maximum value of the Kretschmann curvature invariant is greater 
for a finer resolution and looks divergent in the limit of infinite resolution.
We have also found a similar tendency for the Weyl curvature invariant. 
In this sense, our results also lend support to the formation of 
a naked singularity like in the ST case
with an axially symmetric spindle collapse. 
It should be noted that, even if we did not find an apparent horizon, 
the singularity could be covered by the global event horizon. 
For instance, in Ref.~\cite{Ponce:2010fq}, such possibility is addressed 
for initially stationary configurations of pointlike and singular line sources. 
The results in Ref.~\cite{Ponce:2010fq} indicate the presence of 
a naked singularity when the size of the singular source is large enough 
compared with its mass. 
The event horizon search in the system treated in this paper 
is beyond the scope of this paper and we leave it as a future work.

One remarkable difference from the ST case is that 
the peak position of the Kretschmann invariant always stays inside 
the matter distribution, 
while it is outside the matter distribution near the pole for the ST case. 
The reason of this difference is not quite clear. 
A consistent result with ST is also reported in Ref.~\cite{Yamada:2011br} 
by Yamada and Shinkai.  
They performed a similar simulation to that in ST by using an axisymmetric code 
with a finer resolution. 
Therefore, the lower resolution in ST is unlikely to be the reason. 
One possible reason is 
that there is no exact axisymmetry in our case in contrast with the ST case. 
If it is correct, our result might suggest 
structural instability of 
the singular spacetime suggested by ST. 
Another difference from the ST case is that 
our numerical integration does not break down even when 
the Kretschmann invariant takes a maximum value but goes further 
well beyond this moment. 
We have not found an apparent horizon in the simulation 
starting from the initial situation same as in ST. 
As is shown in the section \ref{subsec:horizon}, 
by using a different initial data set, 
we have found an example in which an apparent horizon is finally formed 
after a certain period of time after the occurrence of 
the maximum value of the peak of the Kretschmann invariant. 

Not only do the analyses of collisionless particle systems
make contributions to the understanding of the theoretical 
aspects of gravitational collapse, but also 
provide a model of gravitational collapse in a 
possible early matter-dominated phase 
of our Universe~\cite{1980PhLB...97..383K,1982SvA....26..391P,2009PhRvD..80f3511A,2012JCAP...09..017A,2013JCAP...05..033A}. 
Similar analyses in cosmological situations may also make 
help to understand the criterion for primordial black hole formation 
in the matter-dominated phase~\cite{Harada:2016mhb}. 
In order to make our setting more realistic for gravitational collapse 
in cosmological situations, we need to choose an appropriate boundary 
condition (e.g., periodic boundary condition as in Refs.~\cite{Yoo:2012jz,Yoo:2013yea}) 
and initial data~\cite{Shibata:1999zs,Harada:2015yda}. 
Gravitational collapse of a collisionless particle system in an expanding universe 
will be reported elsewhere~\cite{Yoo-Harada-Okawa_prep}. 

\section*{Acknowledgements}
We thank K. Nakao, M. Sasaki, M. Siino, S. Inutsuka and H. Shinkai for helpful comments. 
This work was 
supported by JSPS KAKENHI Grant Numbers JP16K17688, JP16H01097 (C.Y.),
JP26400282 (T.H.). 

\appendix

\section{Convergence check}
\begin{figure}[htbp]
\begin{center}
\includegraphics[scale=1]{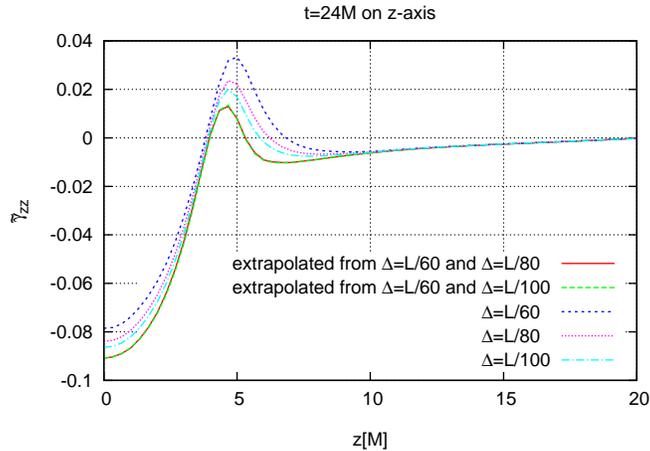}
\caption{$\tilde \gamma_{zz}$ on $z$-axis at $t=24M$ 
is shown for each value of $\Delta$.  
The dependence of the second order convergence is clearly confirmed. 
}
\label{fig:convz}
\end{center}
\end{figure}

In this Appendix, we show a result of the convergence check for 
our numerical code. 
In Fig.~\ref{fig:convz}, we plot the value of 
$\tilde \gamma_{zz}$ on $z$-axis at $t=24M$ for each resolution 
with the following fixed parameters:
\begin{equation}
\frac{L}{M}=20~,~~
\frac{b}{M}=10~,~~
e=0.9~,~~ 
N=125000~,~~
r_{\rm s}=2L/75. 
\end{equation}
Assuming the second order convergence, 
we consider the following resolution dependence of $\tilde \gamma_{zz}$: 
\begin{equation}
\tilde \gamma_{zz}(z;\Delta) =\tilde \gamma_{zz0}(z)+\tilde \gamma_{zz2}(z)\Delta^2
+\mathcal O(\Delta^3), 
\end{equation}
where $\tilde \gamma_{zz0}$ and $\tilde \gamma_{zz2}$ represent 
the true value and second order error, respectively. 
Using $\tilde \gamma_{zz}$ for two different resolutions $\Delta_1$ and $\Delta_2$, 
we can estimate the true value $\tilde \gamma_{zz0}$ with the following extrapolation:
\begin{equation}
\tilde \gamma_{zz0}(z) \simeq \frac{\tilde \gamma_{zz}(z,\Delta_1)\Delta_2^2
-\tilde \gamma_{zz}(z,\Delta_2)\Delta_1^2}{\Delta_2^2-\Delta_1^2}. 
\end{equation}
In Fig.~\ref{fig:convz}, two different extrapolations 
agree with each other. 
This result clearly show the second order convergence of our code with 
fixed number and size of particles.

%

\end{document}